# Understanding Currencies in Video Games: A Review

Amir Reza Asadi
Tehran, Iran mail@amirreza.info

Reza Hemadi
Tehran, Iran
r.hemadi@gmail.com

***Abstract*— This paper presents a review of the status of currencies in video games. The business of video games is a multibilliondollar industry, and its internal economy design is an important field to investigate. In this study, we have distinguished virtual currencies in terms of game mechanics and virtual currency schema, and we have examined 11 games that have used virtual currencies in a significant way and have provided insight for game designers on the internal game economy by showing tangible examples of game mechanics presented in our model.**

## I. Introduction

Gold, Coins, Gems, Points, Action points, or credits, whatever game designers might call them, are part of game mechanics that rule the world in most video games. Advancements in internet technology and smart device hardware has allowed game studios to invest in massively multiplayer online games with realistic graphics and stunning environments. Moreover, designers of these games have designed game mechanics that demand virtual goods; therefore, they presented virtual currencies in the world of their video games. With the players increasing interest in spending real money in exchange for virtual goods, in-game micro-transactions became an expectable feature in video games [7]. At some point, the amount of money spent on in-game currencies reached billions of dollars, and it made governmental monetary authorities cautious about the virtual currency schemes [15], and they began studying the economic aspects of virtual currency and virtual economy. In this study, the game mechanics of virtual currencies in video games will be discussed with regard to game design and economic aspects.

The contributions in this research includes:

1. Research model which combine marketing, game mechanics and monetary aspects of video game that allows game designers to have a comprehensive understanding in designing the economy of the game.

2By examining selected titles of popular video games in each genre, these mentioned aspects were highlighted, and their presence in the game is indicated with tangible examples.

The structure of this paper will be first to provide a research model which encompasses internal economy in-game mechanics and virtual currency schema from monetary aspects, then an analysis of the currencies in video games is provided and is followed by a discussion about characteristics of currencies and conclusion.

## II. Literature Review and related works

In this section, the related works are discussed, and after that, game mechanics, virtual currency, and scheme concepts are reviewed. Video game studies are inter-disciplinary in nature, and understanding the roles of currencies in video games requires even a multidisciplinary effort, so even there are numerous well-researched journals, finding appropriate research is not easy, and using non-academic resources is inevitable, even in academic resources are found related to the research subject, their authors used non-academic resources, so to overcome the theoretical gap, the game designing books were used too.

Yamaguchi [16] has analyzed virtual currency in massive multiplayer online role-playing games, with a focus on the Everquest video game. Everquest is developed by Sony Online Entertainments and was released in 1999. It is one of the first successful 3d MMORPG games, and it had around 440000 players worldwide. The research explored the real-world trading of virtual items and describes that players earn skills based on in-game experiences but reaching high levels requires time allocation, so players who do not have enough time find virtual trading items with real currency rational. In addition, the author argues that although the government does not provide any assurances about the game currencies, players have no concern about the value of the in-game currency because the game creator has the ultimate authority in the virtual world, and it assures the value of the game currency.

Hamari and Lehdonvitra have investigated the demand for creating virtual goods of video games from the marketing point of view.

Crowley [17], in the book 'The Wealth of Virtual Nations,' explains that wealth in video games and how trading game currencies could impact players' actions in video games and their virtual worlds.

For a deeper explanation with the aim of reaching a research model, the role of currencies in-game mechanics and monetary view of game currencies are discussed separately in the below subsections:

### *A. Game Mechanics*

Designing game mechanics is an important part of designing any game, from very basic ones to AAA grade titles. Game designer Richard Rouse [11] offers a pragmatic definition for game mechanics, and he describes it as "what the players are able to do in the game-world, how they do it, and how that leads to a compelling game experience" p.310 and [12] provided a more formal definition on game mechanics: game mechanics are methods invoked by agents, designed for interaction with the game state.

Adams and Dorman [1] considered five different types of game mechanics, including physics, internal economy, progression mechanics, tactical maneuvering, and social interactions. Virtual currency is made for the internal economy and social interaction aspects of game mechanics. In early video games, especially arcade ones, collecting points was featured in video games to create competition in the game, so a user with more collected points would have a better ranking in the leaderboard. After a while, the video game economy got more complex, and players were able to consume their collected points in order to get an extra life. This approach was advanced in role-playing games so players could upgrade their items. In other words, resources are virtual currencies in video games. Resources are "any concept that can be measured numerically," such as money, energy, time, or units. "Anything the player can produce, gather, collect, or destroy is probably a resource of some sort" [1]. According to [1] internal economy can be implemented for these aims:

- Complementing physics: Physics are used to evaluate the player's skills, in most action games, the scoring system complements a rewarding system which lets players spend collected points in purchasing power-ups, for example, in Super Mario Brothers players are able to collect coins to gain extra lives [1].
- Influencing the progression: in these games collection a set of points allows users to get access to new locations or levels.
- Adding strategic dimension: Implementing an internal economy is a good way to introduce a strategic dimension to a game. • Creating large probability spaces: internal game economy creates large probability spaces which increase the game's replay value by offering more options to the player while playing the game than is achievable without economy [p76].

As can be seen, especially in mobile games, virtual currencies have a significant role in-game mechanics and the revenue model of video games, so marketers find their place in the video game design process. The authors of [8] analyzed how video game mechanics in combination with marketing concepts create demand for virtual goods. Based on their research, these patterns can lead to an interest in purchasing virtual items.

- Item Degradation: Virtual items degrade with time or usage, for instance, in World of Warcraft, items may be degraded due to usage in combats or items may disappear from inventory due to their expiration date. Furthermore in some cases, item degradation is related to a consumable quantity in items, meaning the number of times an item can be used is limited, for example a consumable item such as a potion can be used by the player only five times.
- Inconvenient gameplay element: in this case, a game specific need is created, to which a virtual good that addresses the need is offered as a solution.
- Medium of exchanges: In MMOs various currencies are used as medium of exchange in purchases, they are also used as rewards for accomplishments, in addition, credits can be used as a "status indicator" and it makes them a "desirable virtual asset" themselves.
- Inventory mechanics: Inventory lots are limited in number.
- Special occasions have been actively used by virtual world operators to promote virtual item sales.
- Artificial Scarcity: Scarcity is a common strategy in traditional marketing, in video games, scarcity is more commonly achieved by making certain items are difficult to obtain through gameplay. These items are most commonly not purchasable and thus do not make revenue for the game publisher directly. However, users may well be encouraged to spend money on purchasable items that help them reach the rare and desirable items.

### *B. Virtual Currency Schema*

For a thousand years, issuing coins were under the control of governments, but video games and virtual worlds within them started to use their own currencies and even paved the way for cryptocurrencies [13]. Due to the large value of exchanged virtual currencies, monetary organizations started investigating the economic aspects of virtual currencies. The European Central Bank [15] has provided the definition below.

"Virtual currency is a type of unregulated, digital money, which is issued and usually controlled by its developers, and is used and accepted among members of a specific virtual community".

According to their research on connections of virtual currency and real money, they define three types of virtual currencies shown in figure 1, which are described in the following.

1. Closed Virtual Currency: this type of currency has no connection to real money in theory, and users earn it based on their activities. However, if it's possible to transfer virtual currency from one player to another member of the community, exchanging virtual currency with real money is sometimes unavoidable even if it's prohibited by the publisher.
2. Virtual currency with unidirectional flow: This virtual currency can be purchased with a certain exchange rate but it cannot be converted back to real currency, Nintendo points are an example of this kind of currency.
3. Virtual currency with bidirectional flow: Players are able to buy and sell this currency with real money, Linden Dollar which is issued for the game Second life virtual world is an example of this type of currency.

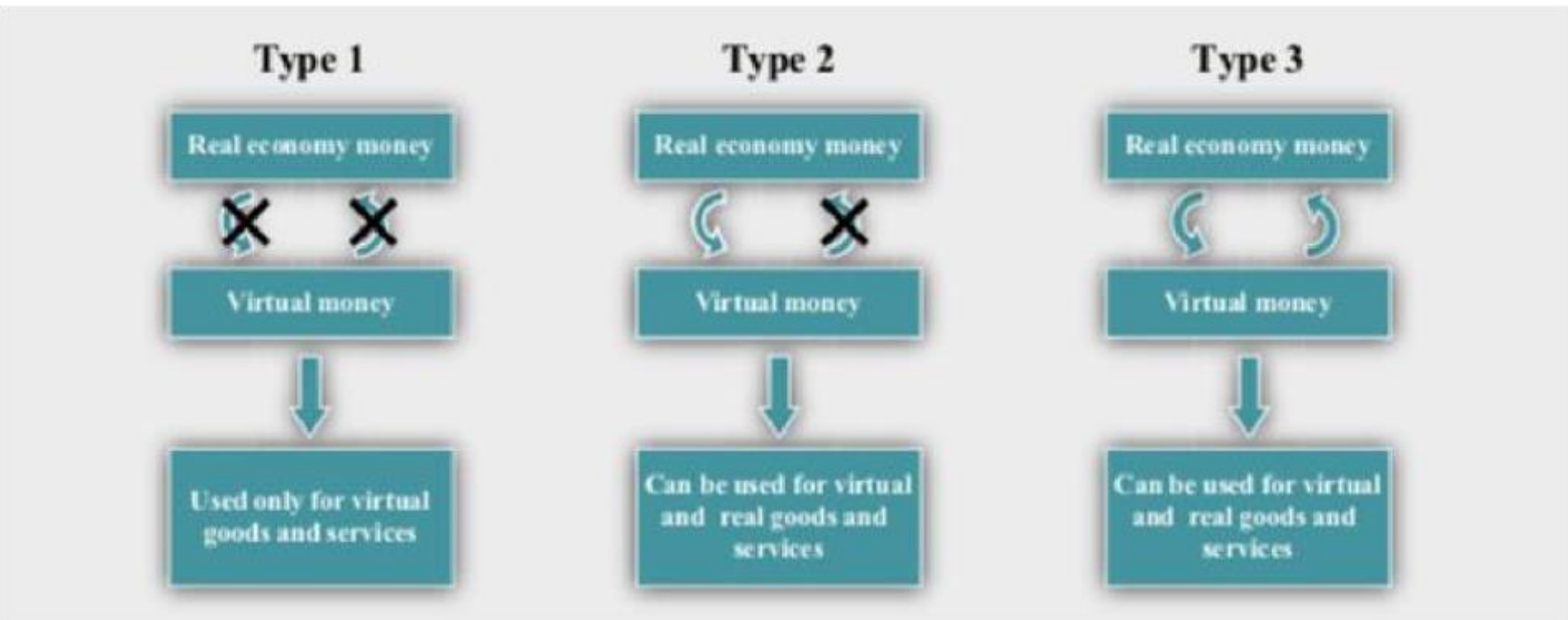

Fig 1. Types of virtual currencies[15]

In regard to works conducted by [15],[8] and [1] the research model has designed which is presented in Fig 2.

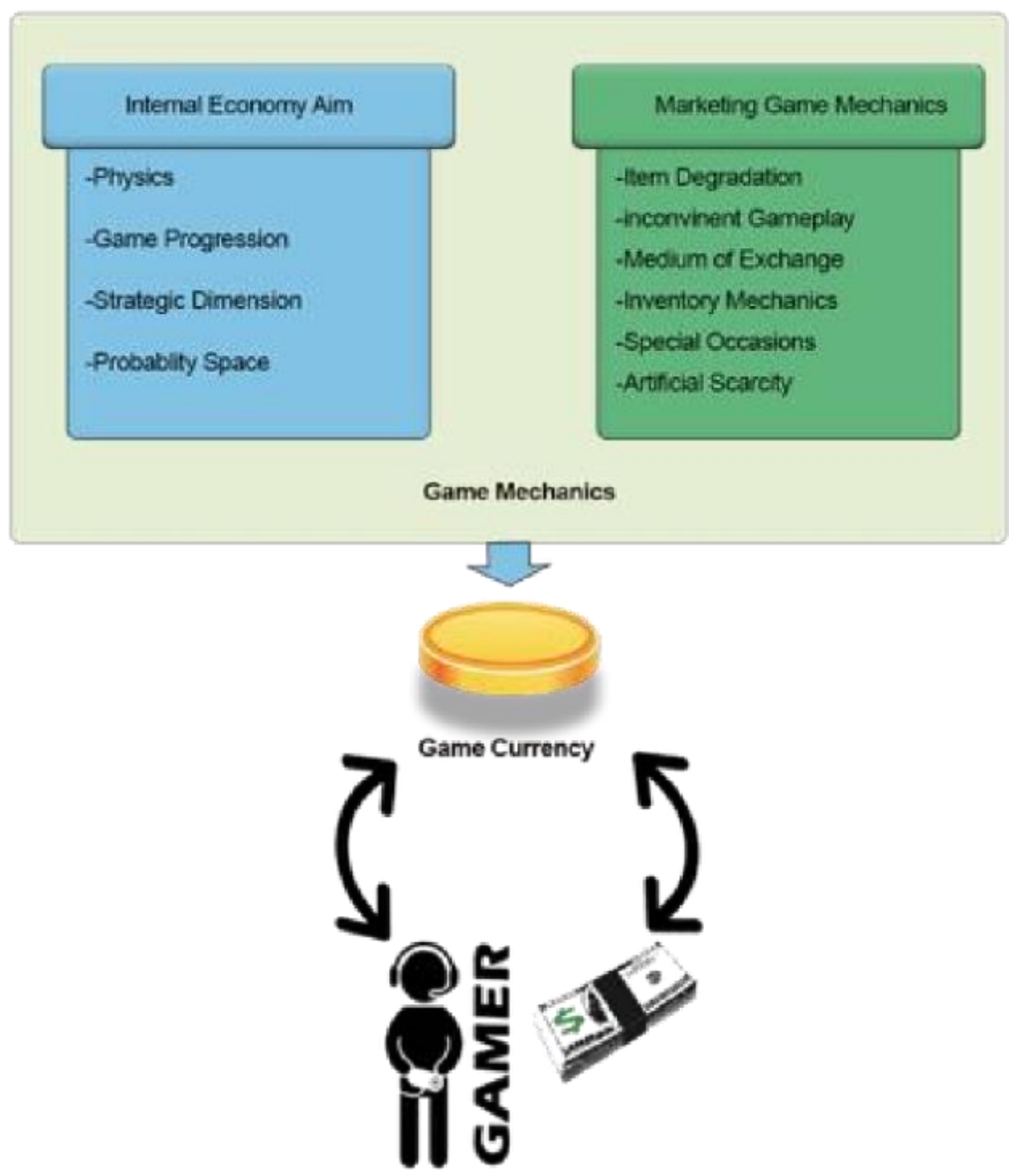

Fig2. Research model.

## III. RESEARCH METHODOLOGY

The research is descriptive and carried out as a desk study and a field assessment. The literature review and analysis of games by playing was designed as a descriptive research to provide baseline information on existing applications of currencies in game design.

To define the domain of video games to investigate in this study, the customers of an online video game shop were asked to mention the list of video games with noticeable use of in-game currencies and resources, also the participants whose age was less than 25 were discarded to include a wider range of video games and to avoid bias in studied games because younger gamers might not have experienced older games. An electronic form was presented to them and their responses were gathered (Fig 3).

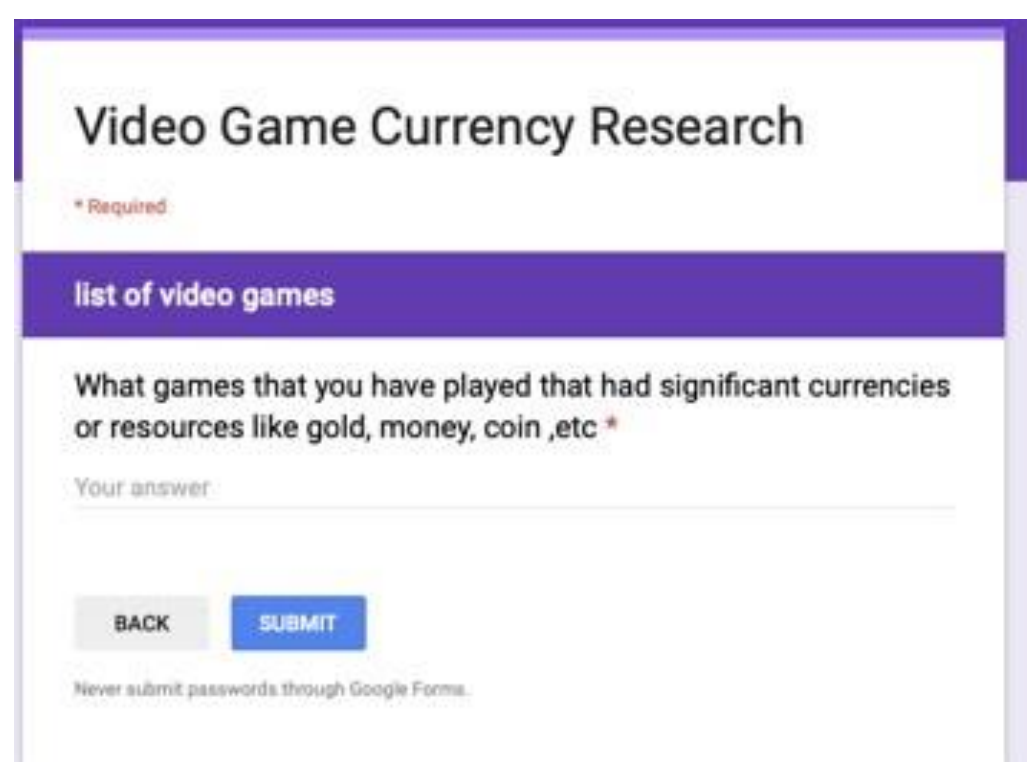

Fig 3

The total of 51 responses were collected but responses of 29 participants were from older gamers, however they were enough to cover different aspects of video game currencies. The results created a list of 33 video games and there were classified by their genres which are gathered in table 1.

| SELECTED GAME FOR DETAILED ANALYSIS | LISTED GAMES BY RESPONDENTS | CLASSIFICATION |
|---|---|---|
| FIFA Series | FIFA ,NBA2K, NHL, Madden, The Crew | SPORT GAMES |
| Rainbow Six Siege | Rainbow Six Siege - Call of Duty Black Ops 3 | ACTION SHOOTER GAME |
| Rise of Nations, Clash of Clans | Rise of Nations, Command and Conquer Generals, Warcraft 3, Age of Mythology, Civilization Series, Clash of Clans, Clash of Kings | STRATEGIC GAMES |
| Fable 3, World of Warcraft | Skyrim, Dragon Age Inquisition, Diablo 3, Fable 3, World of Warcraft | ROLE PLAYING GAMES |
| The Sims and The Sims Freeplay | The Sims, Sim City, Zoo Tycoon, Tropic 5, The Sims Free Play | SIMULATION |
| Minecraft | Minecraft | SANDBOX SURVIVAL |
| SecondLife | Second Life, Avakin Life,IMVU | VIRTUAL WORLD |
| Gangstar Vegas | GTA V, Gangstar Vegas, Mafia 3, Sleeping Dogs, Yakuza | ACTION ADVENTURE SANDBOX |

Table1. Considered games based on received responses.

After that authors played all 33 video games and used MIT game analysis guidelines [] To explore aspects of selected video games, the features and mechanisms of these video games were compared to the research model elements and the authors asked the following questions from themselves during the process of playing these games:

1. What are the resources and currencies in this video game?
2. What are the aims of implementing these currencies in this video game?
3. How game mechanics create demand for currencies in the game?
4. Is it possible to purchase game currencies with real money?
5. Is it possible to sell game currencies and earn real money?

As mentioned in MIT game analysis guidelines, the authors used walkthroughs and tutorials from broadcasting websites because there are no academic walkthroughs for video games. Using walkthroughs helped researchers explore less obvious aspects of the relationship between the internal economy and game mechanics. Because of similarities in the design of game currency in titles in a single genre, certain games of each genre were selected for close inspection, with the exception of The Sims, which offers two different versions for PC and mobile with premium and freemium business models, respectively. Besides that, ammo or health was omitted from inspection because they are the basic types of resources.

## IV. Video Game currencies

This section reviews the recognizable currencies in selected video games.

*1) World of Warcraft*

World of Warcraft is a massive multiplayer online role-playing game released in 2004 by Blizzard Entertainment. Gold is the currency in the game, and players earn it by finishing quests. The game was known to have a restricted system of purchasing gold where in theory, players couldn't buy any Gold with real money, and it was categorized as closed virtual currency. However, in online communities and the grey market, it was traded with real money for prices like 20$ for about 15k to 20k gold coins [2]. In 2015 Blizzard introduced a game time token, which could be bought with real money, and players can exchange this token for gold in the game. Therefore the golds of WoW are unidirectional currency. World of Warcraft used virtual currencies to complement game physics, influencing the progression, adding strategic dimension, and creating large probability spaces. Furthermore, from marketing aspects, the game uses special occasions for discount, item degradation, medium of exchange, and artificial scarcity to increase players' desire to purchase virtual goods.

*2) The Sims PC Version*

The Sims is a life simulation video game series published by Electronic Arts in 2000. It uses Simoleon § as its currency. Simoleons are assigned to the player to spend on any Sim they control. Because The Sims is a simulation of the real world, players can earn money in most ways. They can earn money in real life, for example, by having a regular job. Earned Simoleons can be spent on home improvements, decorations, appliances, clothing, and real state.

From a monetary point of view, Simoleon has no connection to real money and it should be categorized as closed currency.

The game has used Simoleons to create large probability spaces that compel the player to replay the game to explore different possibilities. Also, the game benefits from item degradation mechanics when a Sim loses a property or is rubbed by a thief.

*3) The Sims FreePlay (Mobile)*

The Sims FreePlay is a free version of the Sims video game series, which is a strategic life simulation released in 2011. In addition to Simoleons, the game uses Lifestyle Points, Social Points, and VIP points as currencies.

Earning virtual currencies is almost like the PC version, but it's scarcer and harder to earn by participating in ingame activities. However, it's available for purchase in the game store for a few dollars. Social points can be collected by achieving social goals, but the reward is small compared to the cost of items players can buy with social points. Like Simoleons, Social points can be purchased with real money from the online store.

LifeStyle points are a type of currency used to purchase premium items [14]. Players can earn LifeStyle points by completing a hobby activity but it might be too time consuming so it's easier to buy them with real money from the store. If players buy Simoleons, LifeStyle points, or Social points from the store, they will be rewarded with VIP points. VIP points allow players to overcome inconvenient time-consuming tasks by spending a few of these points. Normally, without spending any VIP Points, players would have to spend a lot of time waiting for some tasks by Sims to finish, which is hardly done by anyone because it takes a long time. However, by spending VIP Points, they can get those tasks done immediately. This is a typical example of using virtual currencies to overcome inconvenient in-game activities.

Because these VIP Points can't be gifted or exchanged among players, this currency falls into type 2 of virtual currencies.

*4) Minecraft*

Minecraft is a sandbox video game developed in Sweden in 2009. In Minecraft, players interact with the game world by placing and breaking different types of 3D blocks, which let players create structures, buildings, and artifacts with their mined blocks [6], so resources play a significant role in this video game and from the game's internal economy view, each of the collectible units is a type of currency. However, the dominant currency in this game is Emerald. Players can earn it by trading with villagers, players can also mine it in the game, but this material is very rare and uncommon. From monetary aspects, emeralds are a closed virtual currency. Emerald and other resources in this game add to the strategic dimension of the game and have a huge impact on probability space, and compels the player to play more and more. Furthermore, Minecraft has adopted item degradation in the game, for instance, axes which are useful tools in the game will lose their heads after long usage, and players should craft new axes..

In 2017 Minecraft got its own purchasable currency called Minecraft Coins. Players can buy Minecraft Coins with real money and spend them for buying texture packs, adventure maps, mini games and more [Stacey Liberatore], providing these items make the probability space infinite, so it can keep the game in the market for many years.

*5) Second Life*

Second life is the largest 3D online virtual world released in 2003. Although the company behind it believes Second Life is not a game and it presents an open-ended experience, it is considered a game in various media and is compared with World of Warcraft and The Sims game series, even though there's no objective or mission to follow in Second Life, and mini-games are a part of endless explorations that users can enjoy.

Second Life residents are able to create their own customizable 3D avatars and travel to various locations and imaginary islands to face new possibilities, and players can meet new people, play games, and interact with objects in this virtual world. The Lindon Dollar L$ is the currency in the Second Life economy, and the player can buy it with real money from the linden exchange. The Linden dollar can be used for buying lands and Second Life's original services. Furthermore, residents of this virtual world can buy or sell services such as education services or business consulting, or virtual goods, including buildings, vehicles artworks, or animation with the Linden Dollar [3]. Players can also exchange their earned Linden dollar with US Dollars, making the Linden dollar a bidirectional virtual currency.

Since Second Life is not limited to in-game design boundaries, analyzing it from a game mechanics perspective is hard, but there are a few points to make. Second Life's economy creates a boundless probability space that allows the company behind it to serve millions of users for many years. Besides, the creator company uses special occasions and inconvenient gameplay elements. For instance, teleporting into the game needs spending Linden Dollar making the game more believable, just like spending money to travel in real life, the player has to spend virtual money to travel in the game, which makes the virtual currency more valuable and makes the player spend real money to buy virtual currency.

In summary, Linden dollar is important because of its bidirectional flow, brilliant feature of earning exchangeable money from virtual experience taking in game currencies to a higher level.

*6) RainbowSix Siege*

RainbowSix Siege is a long-running series that was released by Ubisoft in December 1, 2015. RainbowSix Siege is a tactical first-person shooter that uses “Renown” and “R6 Credits” as currencies. Players earn Renown for almost any activity in the game. However, by spending R6 Credits, players can purchase Renown boosts to earn Renown more rapidly. Using Renown, players can purchase operators, weapons, and weapon skins. Items purchased using Renown complement game physics that evaluate the player’s skills. They also expand the game’s probability spaces and increase the game’s replay value.

R6 Credits on the other hand are optional and can be purchased in game. R6 credits can be used to earn Renown faster by purchasing boosters or to buy rare premium skins. They also can be used to unlock operators instead of using Renown. R6 Credits influence the game progression by offering the players to gain access to extra operators.

RainbowSix currencies are closed since they cannot be exchanged with real money or be transferred to other players.

*7) Fifa 13 – Now*

Fifa is a video game series released by Electronic Arts since 1993 and it allows players to enjoy the simulation of football. By Fifa 13, EA introduced two currencies for Fifa Ultimate team, which is the game’s online mode called Fifa Point and Fifa Coin. Players can purchase Fifa Point packs from online stores and use them to buy game packs. Game packs offer players that can be picked to play in user’s ultimate team, on the other hand, Fifa Coins are not purchasable, and the publisher has prohibited selling and buying Fifa Coins because users should earn Fifa Coins by transferring their soccer players, however in the grey market there are individuals who spend money on Fifa Points then transfer their valuable players in exchange for real money.

From the internal economy point of view, Fifa Coin is categorized as type 1 currency and Fifa Point as type 2, with the gray market following its own path. Buying skilled players influences the game state by complementing the user’s skills in the game, making the game a more exciting experience.

*8) Rise of Nations*

Rise of Nations is a strategy computer game published by Microsoft Game Studios in May 2003. There are six basic resources in the game which allow for the creation of units, buildings, and technologies for a nation. There is no mechanism in the game to buy these resources for real money. Food, Timber, Metal, and Oil are harvested in the game using Citizens. Wealth is gained by expanding territories or by using Caravans to trade between cities. Knowledge is gained using Scholars in the game. These resources form the basis of the strategic dimension of the game and are type 1 closed currencies that can only be gained in the game. Each of the basic resources except for Wealth requires a building and civilian units to gather/harvest them. There’s another limitation in the game, which is the Population limit. It puts a limit on the number of units the player can create. Population limit can be increased by building Cities and thus adds to the strategic value of the game.

*9) Clash of Clans*

Clash of clans is a strategy mobile game released in 2012 by SuperCell. There are four collectible resources in the game: Gems, Gold, Elixir, and Dark Elixir. Gems could be considered the main virtual currency in the game since they can be used to buy other resources like Gold or Elixir. Gems can be bought with real money within the game. The game offers players to collect Gems in very few quantities. Gems could be used to finish in game processes that require time instantly. The way that Clash of Clans compels players to buy and spend gems is to create inconveniences that take a lot of time, like creating units that require some time to complete and offering the player to omit these inconveniences by spending some Gems. Other resources in the game could be mined or be obtained by attacking other players. They are used to create buildings and units and upgrade them. These items give a strategic dimension to the game where players should be careful on how to spend these resources. Currencies in Clash of Clans are type 2 virtual currencies with a unidirectional flow where players can buy gems and packs of different resources, but there’s no way of selling these resources to other players. Dark Elixir is different from other resources in a sense that there’s a building dedicated to spending Dark Elixir in the game and units created using Dark Elixir don’t need other resources to be generated. This somehow makes it more valuable than other mined resources in the game.

Another vital resource in Clash of Clans is the limit on the number of units the player can create, which is called Housing Space. This is an important currency that complements the player’s strategy and skills because it cannot be exchanged for gems or bought with real money. This resource can be obtained by creating special buildings or upgrading them to hold more units. Created units cost depending on their strength and will occupy more housing space if the unit is more effective. This falls into type 1 closed virtual currency.

*10) Fable 3*

Fable 3 is an open-world fantasy role-playing video game that was developed for Xbox 360 and Windows and released in 2010. The currency of Fable is gold, and players can earn gold by working as a blacksmith, woodcutter, pie maker, lute player, and game activities or trading booties and items with in-game- merchants.

The players need gold for spending estates, weapons, clothes, foods, potions, hairstyles, furniture, and tattoo, which means that the currency of the game creating probability spaces and the need for better weapons is an example of using game currencies for game progression. Buying better armors with collected gold is a kind of complementing physics. Furthermore, there is a judgment game mechanic in the storyline of Fable 3 which players should decide how to fill the debt of treasury, and they can choose spending they earned money and receive positive judgments from civilians or increase taxes on civilians and lose public opinion on the kingdom. So game uses currency to influence the game progression, and it increases probability space at the same time.

Game publisher does not sell gold packs to player, and player should earn gold from game activities, but the multiplayer system of Fable 3 allows gamers to gift their inventory items, including golds, to other players, and there is no official ban from the publisher for exchanging golds with real money. However, the loopholes in-game mechanics bypass the artificial scarcity of gold in the game, so the gold in fable 3 is a closed currency formally and informally.

The players can cut wood for infinite time, and cutting the woods have no impact health of the player; and there are infinite woods for cutting, and even more skilled woodcutters can benefit from the gold multiplier. Suppose the game designers had restricted the woodcutting income for players. In that case, it could have a positive impact on the artificial scarcity of the game, which may attract players for using trading golds with other players.

*11) Gangstar Vegas*

Gangstar Vegas is an action-adventure video game developed by Gameloft for iOS and Android with a freemium business model applied. Game currencies are an important part of the freemium business model because these games use artificial scarcity to force players to spend money on virtual content. Gangstar Vegas also used this mechanism. There are four currencies in this game, including Skill Points, Cash, Moviebucks, and Diamonds.

Players can earn cash by finishing missions or selling stolen cars. In addition, there are daily bonuses for logging in to the game every day. Also, players can exchange hard-earned diamonds with cash. So it's a closed virtual currency. Cash can be used to purchase clothes, drones, vehicles, and weapons, so it's very important to create a large probability space.

-Moviebucks: Player can gain Moviebucks by watching advertisements, watching advertisement generate profits for publisher and players can purchase items like cloths with earned moviebucks. It's a closed currency too and it creates a larger probability space. In addition it uses special occasion technique for offering special items like super hero outfits for a limited time.

-SkillPoints: SkillPoints can be earned by doing game activities like driving and shooting, however it's a time taking effort for players to get skillpoints. Skill points are required to level up players' skills and using high level weapons and vehicles. So the game designer used skill point with aims of complementing physics and influencing the progression.

-Diamonds: this is most valuable currency in this game, they can earned through game activates however they are hard to earn and rare and artificial scarcity has been applied for this currency. Diamonds can be exchanged to SkillPoints and Cash and players can purchase diamonds with real money, so diamond is a currency with unidirectional flow.

Although earning cash may be an easy task for some players, a certain group of weapons and vehicles can be purchased only with diamonds, so the diamonds have an impact on game progression, and it is an element for the inconvenience in gameplay.

Last but not least, the inventory mechanism is restricted, and players need to spend skill points to increase the capacity of inventory.

| Game | Currency | Mechanisms | | | | | | | | | |
|---|---|---|---|---|---|---|---|---|---|---|---|
| | | Complementing physics | Progression | Strategic Dimension | Probability Space Creation | Item Degradation | Inconvenient Gameplay | Medium of Exchange | Inventory Mechanic | Special Occasions | Artificial Scarcity |
| **World of Warcraft** | Gold | X | X | X | X | X | X | X | X | X | X |
| | Time Token | | | | | | | X | | | |
| **The Sims PC** | Simoleon | | | | X | X | | X | | | |
| **The Sims FreePlay** | Simoleon | | | | X | | | X | | | |
| | LifeStyle Point | | X | | X | | | | | X | X |
| | Social Point | | X | | X | | | | | | X |
| | VIP Point | | | | X | | X | | | X | X |
| **Minecraft** | Emerald | | | X | X | X | | | | | |
| | Minecraft Coin | | X | | | | | | | | |
| **Rainbow Six Siege** | Renown | X | | | X | | | | | | |
| | R6 Credit | | | | | | | X | | | |
| **FIFA Series** | FIFA Coin | X | | | | | | | | | |
| | FIFA Point | | | | | | | X | | | |
| **Rise of Nations** | Food | | | X | | | | | | | |
| | Timber | | | X | | | | | | | |
| | Metal | | | X | | | | | | | |
| | Wealth | | | X | | | | | | | |
| | Knowledge | | | X | | | | | | | |
| | Population Limit | | | X | | | | | | | |
| **CLASH OF CLANS** | Gem | | | X | | X | X | X | | X | |
| | Gold | | | X | | | | | | X | |
| | Elixir | | | X | | | | | | X | |
| | Dark Elixir | | | X | | | | | | X | |
| **Fable 3** | Gold | X | X | | X | | | | | | |
| **Gangstar Vegas** | Cash | | | | X | | | | | X | |
| | Moviebucks | | | | X | | | | | | |
| | SkillPoints | X | X | | | | X | | | | |
| | Diamonds | | X | | X | | | X | | | X |

Table2. To sum up, the summary of inspected currencies and mechanisms that supported these currencies are gathered.

## V. Conclusion

This study has presented an initial effort for understanding the role of currencies in video games from game mechanics and a monetary point of view. The research model and tangible examples of its dimensions in successful video games can be really helpful for game designers because it reveals the industry standards of game currencies. The research model helps game developers to consider aspects like monetary and marketing, which might be far from their educational backgrounds.

Although this study considered game mechanics behind currencies of commercial video games, there might be game mechanisms that aren't considered by the research model, so conducting interviews and content analysis in future work can expand the current model. In addition, the relation between narration and demand for game currencies should be discussed in future works. Furthermore, considering the fact that game mechanics and monetary aspects are interconnected in video games, this paper recommends effective game mechanics that utilize virtual currency to game designers.